\documentclass[10pt,prl,twocolumn,superscriptaddress]{revtex4}

\pdfoutput=1
\usepackage{appendix}
\usepackage{graphicx}		
\usepackage{amssymb}
\usepackage{times}
\usepackage{amsmath}
\usepackage{amsfonts}
\usepackage{txfonts}
\usepackage{color}			
\usepackage{natbib}
\usepackage{units}
\usepackage{textcomp,calrsfs}
\usepackage{mathrsfs}
\usepackage{calrsfs}
\usepackage{ifsym}
\usepackage{bm}
\usepackage{relsize}
\usepackage[stable]{footmisc}

\newcommand{\braket}[1]{\langle#1\rangle}
\newcommand{\eqn}[1]{Eq.~(\ref{#1})}
\newcommand{\fig}[1]{Fig.~\ref{#1}}

\begin{document}
\title{Minimum requirements for feedback enhanced force sensing}
\author{Glen I. Harris} \affiliation{Centre for Engineered Quantum Systems, University of Queensland, St Lucia, Queensland 4072, Australia}
\author{David L. McAuslan} \affiliation{Centre for Engineered Quantum Systems, University of Queensland, St Lucia, Queensland 4072, Australia}
\author{Thomas M. Stace} \affiliation{Centre for Engineered Quantum Systems, University of Queensland, St Lucia, Queensland 4072, Australia}
\author{Andrew C. Doherty} \affiliation{Centre for Engineered Quantum Systems, University of Sydney, 2006, Australia}
\author{Warwick P. Bowen} \affiliation{Centre for Engineered Quantum Systems, University of Queensland, St Lucia, Queensland 4072, Australia}


\begin{abstract}

The problem of estimating an unknown force driving a linear oscillator is revisited. When using linear measurement, feedback is often cited as a mechanism to enhance bandwidth or sensitivity. We show that as long as the oscillator dynamics are known, there exists a real-time estimation strategy that reproduces the same measurement record as any arbitrary feedback protocol. Consequently some form of nonlinearity is required to gain any advantage beyond estimation alone. This result holds true in both quantum and classical systems, with non-stationary forces and feedback, and in the general case of non-Gaussian and correlated noise. Recently, feedback enhanced incoherent force sensing has been demonstrated [Nat. Nano. \textbf{7}, 509 (2012)], with the enhancement attributed to a feedback induced modification of the mechanical susceptibility. As a proof-of-principle we experimentally reproduce this result through straightforward filtering.
 
\end{abstract}

\maketitle


Micro and nano-mechanical oscillators are capable of ultra-sensitive force measurement, allowing precision spin, charge, acceleration, and field sensing~\cite{Rugar04_Nat, Forstner12_PRL, Cleland98_Nat, Mamin01_APL}. It is well known that linear feedback control can improve the performance of nonlinear mechanical sensors~\cite{Harris12_PRA, Leang07_IEEE}. For example, in non-contact atomic force microscopy linear feedback is commonly used to stabilise the tip-surface separation, thereby avoiding collisions and suppressing frequency drifts due to short range forces such as van der Waals forces~\cite{Binnig86_PRL, Lee02_PRB}. Since feedback control modifies the response of an oscillator to environmental forces it also appears attractive as a technique to enhance the performance of {\it linear} sensors. For instance feedback cooling allows the suppression of thermal noise~\cite{Mertz93_APL,Cohadon99}, while feedback tuning of the spring constant can provide increased mechanical response at the signal frequency~\cite{LIGO09_NJP}. However, such precision enhancement is prohibited for linear processes with stationary linear feedback and uncorrelated Gaussian noise by the well-known principle of neutrality in linear control theory, which states that the accuracy with which the oscillator position can be determined is independent of feedback~\cite{Jacobs74, Patchell71_IJC}.

Non-stationary processes, non-Gaussian noise and non-linear estimation strategies are each found in a range of linear oscillator-based force sensors. Stroboscopic measurement of impulse forces~\cite{Vitali01_PRA}, and variance estimation of incoherent forces as applied in bolometry~\cite{ Levin65_IEEE, VanTrees71_Book}, are two relevant examples. Linear feedback cooling has been proposed as a means to enhance precision in both cases~\cite{Vitali01_PRA, Gavartin12_NatNano}, and experimentally demonstrated in the latter~\cite{Gavartin12_NatNano}. However, neither proposal identifies an optimal estimation strategy. This leaves unresolved the important question of whether the same, or improved, sensitivity enhancement might be achieved without feedback by applying a better estimation strategy.

\begin{figure}[ht!]
\begin{center}
\includegraphics[width=8cm]{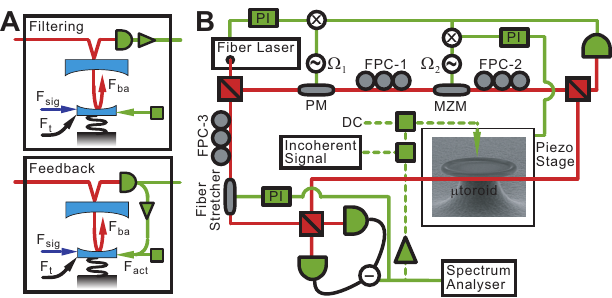}
\caption{(Color online). {\bf A} Conceptual diagram comparing optomechanical force sensing with feedback and filtering {\bf B} Experimental schematic. Red (dark grey): fiber interferometer; solid green (light gray): electrical components for feedback stabilization; dashed green: electrical signal applied to the microtoroid for application of the incoherent force and feedback. FPC: fiber polarization controller. BPF: bandpass filter. MZM: Mach-Zender Modulator. PM: phase modulator. PI: proportional-integral controller}
\label{fig1}
\end{center}
\end{figure}

Here we present a straightforward theoretical approach which shows that, even in the presence of non-Gaussian or correlated noise and non-stationary processes, a real-time estimation strategy always exists that reproduces the same measurement record as any arbitrary linear feedback protocol (see \fig{fig1}A). The theory applies to both quantum and classical oscillators, and to the intrinsically non-linear problem of variance estimation~\cite{Gavartin12_NatNano}. It ultimately provides a clear set of minimum requirements for feedback to provide any advantage to force sensing over that possible with estimation alone. Essentially, some form of nonlinearity is required in the physical system, which may arise from the measurement process, feedback loop, signal, or from the oscillator itself. When no non-linearities are exhibited, the theory yields a filter which allows the estimate that would be obtained with feedback to be determined causally from the measurement record without feedback. This precludes the possibility of any additional sensitivity enhancement from feedback in either of the examples discussed above~\cite{Vitali01_PRA,Gavartin12_NatNano}, or indeed feedback improved bandwidth in linear force sensors~\cite{Miao12_NJP}.

To validate the theoretical results, we experimentally reproduce the effect of feedback enhanced sensitivity achieved in Ref.~\cite{Gavartin12_NatNano}, but replacing feedback with causal filtering. This demonstrates that the sensitivity enhancement achieved in that experiment was not due to a feedback-induced change in coupling of the oscillator to its the environment, as is suggested in Ref.~\cite{Gavartin12_NatNano}. Rather it  arises from an intrinsic bias in the estimator used towards low quality, and thus feedback cooled, oscillators. By clearly demarcating the circumstances in which feedback may advantage force sensing over estimation alone, our results both clarify a significant ambiguity in the optomechanics and force sensing communities, and contribute towards simplifying experimental implementations of ultra-precise force sensing with linear oscillators.


The evolution of a mechanical oscillator can be described, in both classical and quantum regimes~\cite{Ludwig08_NJP} by the equation of motion
\begin{equation}
m \left [ \ddot{x} +  \gamma \dot{x} + \Omega_m^2 x \right ] \!=\! F_{\rm m}(t,x) + F_{\rm act}(t,\tilde{x}) 
\label{eqmot}
\end{equation}
where $m$, $\gamma$, and $\Omega_m$ are the mass, damping rate, and resonance frequency, respectively. For compactness the combined force $F_{\rm m}(t,x)=F_{\rm T}(t,x) + F_{\rm s}(t,x) + F_{\rm ba}(t,x)$ is used, where $F_{\rm T}(t,x)$ is a thermomechanical force due to the coupling of the oscillator to its environment, $F_{\rm s}(t,x)$ is the signal force and $F_{\rm ba}(t,x)$ is a backaction force due to the act of measurement. $F_{\rm act}(t,\tilde{x})$ is an actuation force used for feedback where $\tilde{x}(t) = x(t) + N(t,x)$ is the instantaneous measurement record of the oscillator position $x$, and $N(t,x)$ is the measurement noise which maybe correlated to the backaction noise. In general, the forces and measurement noise can all have non-stationary dynamics, non-Gaussian noise and non-linear dependence on the oscillator position. 
They can each be linearized by Taylor expanding around the mean position of the oscillator $\bar{x}$ and retaining only zeroth and first order terms (See Supplementary Information). The zeroth order terms are independent of fluctuations in the oscillator position, forcing, and measurement noise; and only serve to deterministically shift the mean position of the oscillator. The first order terms are each linearly dependent on only one source of fluctuation; and either act to modify the mechanical susceptibility or introduce incoherent driving.
Higher order terms introduce nonlinearities and instabilities which can give rise to detrimental effects such as saturation and nonlinear dissipation~\cite{Harris12_PRA, Eichler11_NatNano}. By neglecting these higher order terms, we restrict the analysis to the most general linear oscillator experiencing linear feedback. It is important to note that the deterministic shift in mean oscillator position due to the zeroth order terms can affect the force sensitivity; for example, by shifting a cavity optomechanical system onto optical resonance. However, since this is deterministic, and known {\it a priori}, an equivalent displacement may be made to the oscillator without feedback by applying a known external force, as depicted in \fig{fig1}A (see Supplementary Information for details).

To simplify the analysis and present results most relevant to our experiments, in the main text of the Letter we consider the common scenario where the mechanical oscillator's susceptibility is only modified by the feedback force, and therefore drop the first order susceptibility modifying terms in the other forces. We further assume that all processes involved are stationary. These specific assumptions are not necessary for our conclusions, which hold for the most general linear case, including non-stationary processes and first order terms (see Supplementary Information). Under these assumptions, the combined force is given by $F_{\rm m}(t,x)=F_{\rm m}(t,\bar{x})$ and the feedback force is
\begin{eqnarray}
F_{\rm act}(t,\tilde{x}) &=& \int_{-\infty}^{t} g(t-\tau)\tilde{x}(\tau)d\tau \label{F_act1} 
\end{eqnarray}
where $\tilde{x}(\tau)=x(\tau)-N(t,\bar{x})$ and $g(t-\tau)$ is the stationary feedback kernel describing the filter applied to the measurement record. Enforcing causality, namely $g(t-\tau)=0$ for $\tau>t$, simplifies \eqn{F_act1} into the convolution $F_{\rm act}(t,\tilde{x})=g(t)\ast \tilde{x}(t)$. Substitution into \eqn{eqmot} and Fourier transforming then gives
\begin{eqnarray}
x(\Omega) &=& \chi(\Omega)\left[F_{\rm m}(\Omega,\bar{x})+g(\Omega)\tilde{x}(\Omega) \right]
\label{feed}
\end{eqnarray}
where $\chi(\Omega)^{-1} \!=\! m\left[\Omega_m^2 - \Omega^2 + i \gamma \Omega\right]$ is the intrinsic mechanical susceptibility
\footnote{Strictly speaking the Fourier transform can only be taken if the energy of the system is finite. Consequently, unstable scenarios such as regenerative amplification~\cite{Kippenberg05, Taylor12_OptExp}, where the oscillator energy grows exponentially over all time, are not covered by the analysis. However, even in unstable situations, in the physically realistic case where the measurement is performed over a finite time interval the energy does remain finite, and the Fourier transform can be applied.}.
The oscillator position without feedback can be simply obtained by omitting the feedback force $g(\Omega)\tilde{x}(\Omega)$ 
from \eqn{feed}, $x_{0}(\Omega) = \chi(\Omega)F_{\rm m}(\Omega,\bar{x}_{0})$, where the subscript $0$ is used to distinguish from the feedback case.

As discussed earlier, an external force may be applied to equate the mean positions of the oscillator with and without feedback. In the case of cavity optomechanics this amounts to ensuring identical cavity detunings when experiments are initiated. With $\bar{x} = \bar{x}_{0}$, the common forcing terms with and without feedback $F_{\rm m}(\Omega,\bar{x})$ and $F_{\rm m}(\Omega,\bar{x}_{0})$ are identical. Substituting for $x(\Omega)$ and $x_{0}(\Omega)$ in terms of their respective measurement records (eg. $x(\Omega) = \tilde{x}(\Omega) - N(\Omega,\bar{x})$) then gives a completely deterministic equation relating the time domain measurement records that would be achieved with and without feedback
\begin{equation}
 \tilde{x}(\Omega) \!=\! \left [ \frac{1}{1-\chi(\Omega)g(\Omega)} \right ]  \tilde{x}_{0}(\Omega)   \!=\! h(\Omega)  \tilde{x}_{0}(\Omega)  
\label{eq:filter}
\end{equation}
where $h(\Omega)=\chi^{\prime}/\chi$ is the ratio of the modified mechanical susceptibility $\chi^{\prime}$ to the intrinsic mechanical susceptibility. Therefore the exact position record that would be obtained using stationary linear feedback can be retrieved straightfowardly by applying the filter $h(\Omega)$ to the position record without feedback. This precludes enhancements of both sensitivity~\cite{Gavartin12_NatNano} and bandwidth~\cite{Miao12_NJP} beyond that achievable with estimation alone. Since no constraints are placed on the statistics of the driving forces or measurement noise, this result is valid even for non-Gaussian noise or if correlations exist between measurement and process noise, such as those induced by quantum backaction. Furthermore, since it applies directly to the measurement records, rather than a specific parameter estimation process based on them, it holds for both linear and non-linear estimation processes.
We show in the supplementary information it can be generalised to include linear non-stationary forcing and feedback as well as modifications to the mechanical susceptibility due to effects such as optomechanical dynamical backaction~\cite{Arcizet06, Kippenberg05, Cohadon99}. In this case, the required filter is more complex and is, in general, non-stationary, but remains causal. Our results are valid in both the quantum and classical regime and rigorously prove that no force sensing advantage is provided by linear feedback onto a linear oscillator with known dynamics. Consequently, nonlinearities are a minimum prerequisite for feedback to improve force sensing beyond estimation alone.


Recently, enhanced incoherent force sensing was experimentally demonstrated~\cite{Gavartin12_NatNano} by stationary feedback cooling of a linear oscillator. However, as shown here, no sensitivity enhancement is obtained from this method over estimation alone. The exact filter equivalent to the feedback cooling used in Ref.~\cite{Gavartin12_NatNano} is obtained by substituting $g(\Omega) \!=\! -i \rm m \gamma \Omega g_{\rm f}$ into \eqn{eq:filter}, where $g_{\rm f}$ represents the filter's unitless gain. This filter, denoted $h_{\rm c}(\Omega)$, effects the causal map $\tilde{x}_{0}\mapsto \tilde{x}$. We demonstrate this experimentally here in a similar system to that of Ref.~\cite{Gavartin12_NatNano} consisting of a microtoroidal cavity optomechanical system. An intrinsic mechanical mode of the microtoroid is used to transduce an incoherent electrostatic gradient force applied by a nearby electrode~\cite{Lee10}. A whispering gallery optical mode of the microtoroid is used to read out the mechanical motion and thereby determine the variance of the incoherent force.

Our experimental setup is shown in \fig{fig1}B. A shot-noise limited fiber laser at 1550 nm was evanescently coupled into the whispering gallery mode of the microtoroid using a tapered optical fiber. The microtoroid had major and minor diameters of $60~\mu\rm m$ and $6~\mu\rm m$ respectively with a $26~\mu \rm m$ undercut. The mechanical motion of the microtoroid, which induces phase fluctuations on the transmitted light, was detected interferometrically by beating with a bright $3.5~\rm mW$ optical phase reference followed by shot-noise limited homodyne detection. We actively stabilize the toroid-taper separation using an amplitude modulation technique~\cite{Chow12_OptExp} that maintains a constant coupling rate into the optical cavity. Pound-Drever-Hall locking was used to lock the laser frequency to the optical resonance, which had an intrinsic quality factor of $\rm Q_{0} \!=\! 2.6\times 10^{7}$. A 50/50 tap-off after the microtoroid was used to detect the cavity transformed amplitude and phase modulation, providing the error signal for the optical frequency and taper-toroid separation locks. The interferometer was locked midfringe via a piezo actuated fiber stretcher that precisely controls the optical path length in one arm.

\begin{figure}[ht]
\begin{center}
\includegraphics[width=8cm]{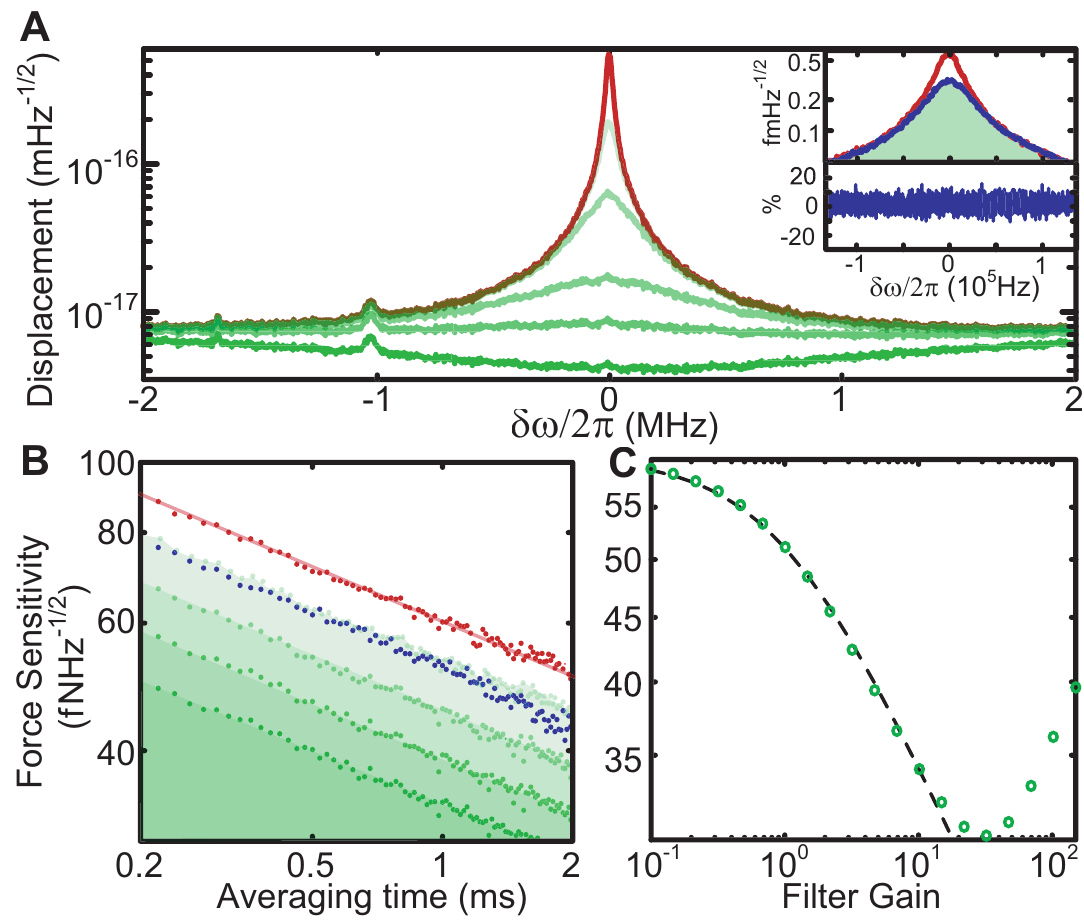}
\caption{(Color online). {\bf A} Displacement spectrum at room temperature (red) and with filtering (green) at gain $2$,$8$,$34$,$72$ and $150$. Inset top: displacement spectrum with feedback cooling (blue) and filtering (green) with gain $1$. Inset bottom: percentage difference between filtered and cooled spectrum with equivalent gain. {\bf B} Force sensitivity as a function of averaging duration for thermal (red points), feedback cooled (blue points) and filtered spectra (green points) at gain $1$,$2.4$,$5$ and $10$. Red line: fit to inverse quartic dependence of the force sensitivity without filtering or feedback. {\bf C} Force sensitivity versus filter gain after $1~\rm ms$ of averaging (green circles) showing good agreement to theory (dashed line).}
\label{fig2}
\end{center}
\end{figure}

The measurement record is acquired from the homodyne signal by electronic lock-in detection where demodulation of the photocurrent at the mechanical resonance frequency allows real time measurement of the slowly evolving quadratures of motion, denoted $I(t)$ and $Q(t)$ where $x(t)\!=\!I(t)\cos(\Omega_{m}t) + Q(t)\sin(\Omega_{m}t)$. Fourier analysis reveals a mechanical power spectra with peaks corresponding to microtoroid mechanical resonances. Fig.~\ref{fig2}A (red) shows the room temperature Brownian motion of a mechanical mode with a signal-to-noise ratio of $37~\rm dB$ and a fundamental frequency, damping rate and effective mass of $\Omega_{\rm m}\!=\!40.33~\rm MHz$, $\gamma \!=\!23~\rm kHz$ and $m_{\rm eff}\!=\! 0.6~\mu\rm g$ respectively. The absolute mechanical displacement amplitude was calibrated via the optical response to a known reference phase modulation~\cite{Schliesser08}.

As shown earlier, applying the filter $h_{\rm c}(\Omega)$ to the measurement record without feedback should retrieve an identical measurement record to that obtained with feedback. Applying this filter to the measurement record it is possible to mimic feedback cooling as shown in \fig{fig2}A (green) where the filter gain $g_{\rm f}$ is varied from $2$ to $150$. Extending the gain beyond $g_{\rm f}>20$ the mechanical spectrum inverts and exhibits squashing, a well known characteristic of high gain feedback cooling~\cite{Lee10}. To confirm the equivalence of the measurement record obtained via feedback and filtering we implement feedback cooling by applying the homodyne photocurrent to the toroid through an electrode which generates strong electrical actuation of the mechanical motion through electrical gradient forces~\cite{McRae10_PRA}. This allows the mechanical mode to be cooled from room temperature by a factor of 2. The upper inset in Fig~\ref{fig2}A shows feedback cooling (blue) and equivalent gain filtering (shaded green); with the fractional difference between the feedback and filtering spectra showing no statistically significant difference (lower inset).

An estimate of the variance of an incoherent force applied to an oscillator may be obtained by determining the oscillator's energy \cite{Gavartin12_NatNano}. After averaging time, $\tau$, the estimate of the energy is given by $E_{\tau} \!=\! 1/\tau \int_{0}^{\tau}dt I(t)^2 + Q(t)^2$. To calculate the ensemble average  $\braket{E_{\tau}}$ and the standard deviation $\sigma_{\rm E}(\tau)$ of the energy estimate multiple independent measurements are made for each $\tau$. Following Ref.~\cite{Gavartin12_NatNano} the energy estimate can be translated into an estimate of the magnitude of the force with a sensitivity given by $\delta F(\tau)^{2} \!=\! \sigma_{E}(\tau)/\int_{0}^{\infty}d\Omega |\chi^{\prime}(\Omega)|^{2}$.
%
%
It is important to note that this estimation process is not necessarily optimal. In the case where $\tau>1/\gamma$ the force sensitivity scales as $(\gamma \tau)^{-1/4}$  which appears to motivate the use of feedback cooling to increase the mechanical decay rate $\gamma$~\cite{Gavartin12_NatNano}.

Figure~\ref{fig2}B (red points) shows the inverse power-law dependence of the force sensitivity on averaging duration, $\tau$, for our experiments with only thermal driving. As predicted by our theory, by applying the filter $h_{\rm c}(\Omega)$ to the thermal data it is possible to enhance the force sensitivity in the same way as feedback cooling. This is shown in \fig{fig2}B (green) where increasing the filter gain, $g_{\rm f}$, provides a clear improvement in sensitivity while consistently maintaining the predicted power-law dependence with respect to averaging time. Figure~\ref{fig2}C (circles) shows the force sensitivity as a function of filter gain taken for a fixed averaging duration of $\tau=1~\rm ms$. For gains below $g_{\rm f}=20$ the measured sensitivity agrees with the theoretical fit. At higher gains it is degraded due to squashing of the mechanical power spectrum which suppresses the signal and allows shot noise to dominate. 

\begin{figure}[ht]
\begin{center}
\includegraphics[width=8cm]{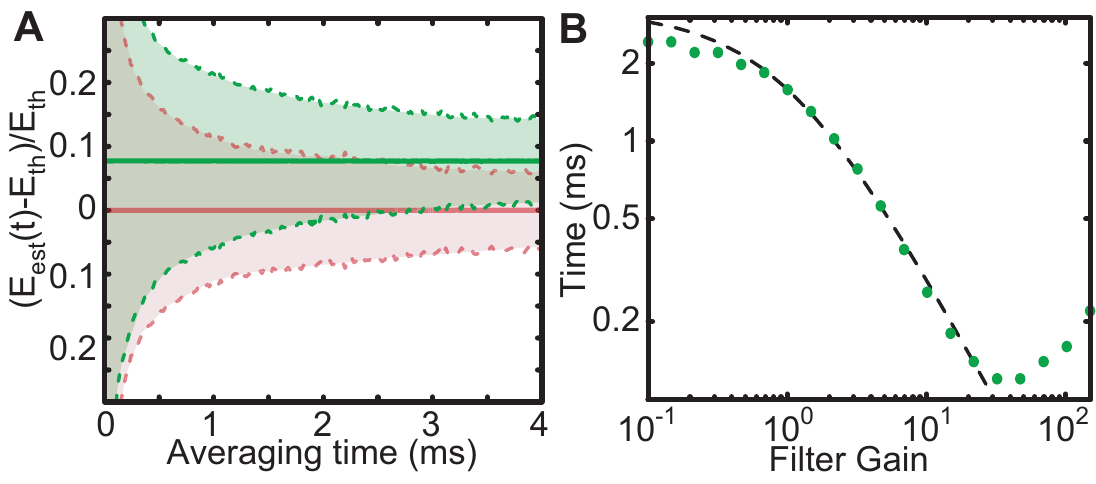}
\caption{(Color online). {\bf A} Normalised energy estimate at room temperature (red), and with addition of incoherent driving (green). Solid lines: mean; dashed lines: one standard deviation error bounds. {\bf B} Averaging time required to resolve incoherent signal versus filter gain (green circles). Dashed line: theory}
\label{fig3}
\end{center}
\end{figure}


To demonstrate the improvement in sensitivity achievable via filtering we apply a small incoherent electrostatic gradient force to the microtoroid with a magnitude of approximately $7$\% of the thermal energy. The ability to resolve this force against the thermal noise depends on the averaging time. Only when the standard deviation of the energy estimate is smaller than the strength of the signal can the incoherent force be resolved. The convergence of the thermal energy estimate with increasing averaging time is shown in \fig{fig3}A (red). With the addition of the incoherent signal the ensemble average is increased without affecting the error bounds as shown in \fig{fig3}B (green). At $~3\rm ms$ the error becomes comparable to the energy separation and the applied incoherent force is resolved. If the filter $h_{\rm c}(\Omega)$ is applied the force sensitivity is improved and the time taken to resolve the applied force decreases as shown in \fig{fig3}B. The required averaging time decreases as the estimation gain is increased in good agreement with theory. At gains $g_{f}>20$ the averaging time increases again due to inversion of the mechanical spectrum and suppression of the signal relative to shot noise. The inflection point in \fig{fig3}B and \fig{fig2}C shows that even though thermal noise dominates shot noise by orders of magnitude at the peak of the mechanical susceptibility, in incoherent force sensing it is shot noise that determines the ultimate sensitivity limit.


The experiments presented here show that feedback and filtering allow equivalent enhancement in incoherent force sensitivity. In this context, the results of Ref.~\cite{Gavartin12_NatNano}, can be naturally understood as a consequence of an intrinsic bias that arises when using the oscillator energy to perform incoherent force estimation. Near-resonant spectral components of the incoherent force drive the oscillator more strongly, and are therefore over-represented in the measurement. As a result, even though feedback only applies a reversible transformation to the measurement record, the force sensitivity appears to improve with increasing oscillator linewidth as the estimation becomes more balanced. The stroboscopic feedback enhanced force sensing scenario proposed in \cite{Vitali01_PRA} is similarly biased. In that case, measurements prior to application of the signal force are used to pre-cool the oscillator. This improves its initial localization in phase space and, thereby, the capacity to resolve displacements due to external forces. However, the existence of this prior measurement record is not taken into account when calculating the sensitivity of measurements without feedback. Filtering it appropriately allows equivalent localisation to feedback cooling, though offset from the origin, and achieves identical sensitivity. These examples illustrate the main result of this Letter that, for a linear oscillator, any sensitivity enhancement arises not through the action of feedback, but rather through measurement and estimation alone.

In summary, we have theoretically shown that for linear oscillators neither stationary nor non-stationary linear feedback provide any force sensing enhancement over that possible with estimation alone. This holds true in both the quantum and classical regime and in the presence of non-Gaussian noise. As a demonstration we have shown experimentally that detection of a stationary incoherent force can be enhanced equally effectively via estimation as with feedback cooling. These results may have broad relevance to many scientific and engineering communities, particularly those associated with mechanical sensors limited by thermomechanical noise. {\it Acknowledgments}: This research was funded by the Australian Research Council Centre of Excellence CE110001013 and Discovery Project DP0987146. Device fabrication was undertaken within the Queensland Node of the Australian Nanofabrication Facility.

\bibliography{ForceSensingExpBIB}{}
\clearpage

\appendix
\section{Supplementary Information}

Here we show that the theoretical result presented in the main text of the Letter can be generalised to include linear non-stationary forcing and feedback as well as modifications to the mechanical susceptibility due to effects such as optomechanical dynamical backaction~\cite{Arcizet06, Kippenberg05, Cohadon99}. Due to non-stationarity, the forces cannot be expressed as a simple convolution and hence require a more complex theoretical approach.

The generalized optomechanical system considered here is shown schematically in \fig{fig1}A. The evolution of the mechanical oscillator can be described, in both classical and quantum regimes~\cite{Ludwig08_NJP} by the following equation of motion
\begin{equation}
m \left [ \ddot{x} +  \gamma \dot{x} + \Omega_m^2 x \right ] \!=\! F_{\rm m}(t,x) + F_{\rm act}(t,\tilde{x}) \label{eqmot_Supp}
\end{equation}
where $m$, $\gamma$, and $\Omega_m$ are the mass, damping rate, and resonance frequency of the mechanical oscillator, respectively. For compactness the combined force $F_{\rm m}=F_{\rm T}(t,x) + F_{\rm s}(t,x) + F_{\rm ba}(t,x)$ is used, where $F_{\rm T}(t,x)$ is a thermomechanical force due to the coupling of the oscillator to its environment, $F_{\rm s}(t,x)$ is a general signal force and $F_{\rm ba}(t,x)$ is a backaction force due to the act of measurement. $F_{\rm act}(t,\tilde{x})$ is a linear actuation force where $\tilde{x}(t) = x(t) + N(t,x)$ is the instantaneous measurement record of the oscillator position $x$, and $N(t,x)$ the measurement noise which maybe correlated to the backaction noise. In general, the forces and measurement noise can all have non-stationary dynamics, non-Gaussian noise and non-linear dependence on the oscillator position.

\begin{figure}[ht]
\begin{center}
\includegraphics[width=8.5cm]{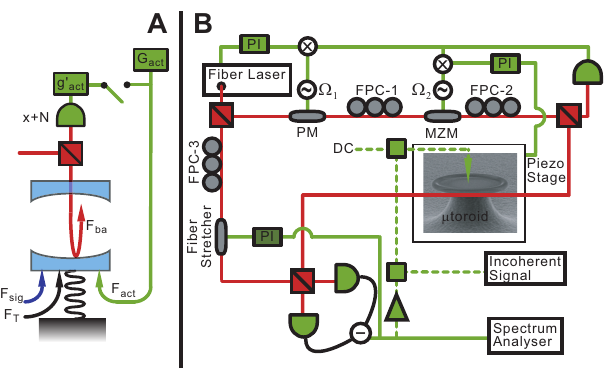}
\caption{(Color online). {\bf A} General theoretical schematic outlining the emergence of forces and feedback. {\bf B} Experimental schematic. Red (dark grey) indicates the fiber based interferometer; Solid green (light gray) indicate the electrical components for feedback stabilization and dashed green represent the electrical signal applied to the microtoroid for feedback and application of incoherent force. FPC: Fiber polarization controller. BPF: Bandpass filter. MZM: Mach-Zender Modulator. PM: Phase modulator. PI: Proportional-integral controller}
\label{fig1_Supp}
\end{center}
\end{figure}

The combined thermal, signal, and backaction force may be expanded without loss of generality, as
\begin{eqnarray}
F_{\rm m}(t,x)&=&\int_{0}^{t}d\tau g_{\rm m}(t,\tau,x) + n_{\rm m}(t,\tau,x)\xi_{\rm m}(\tau) \label{F_fb1_Supp} \\
					&=& \int_{0}^{t}d\tau g_{\rm m}(t,\tau,\bar{x}) + g'_{\rm m}(t,\tau,\bar{x})(x(\tau)-\bar{x}) \nonumber\\
					&&\hspace{5mm} + n_{\rm m}(t,\tau,\bar{x})\xi_{\rm m}(\tau) \label{F_fb_Supp}\\
					&=& \int_{0}^{t}d\tau G_{\rm m}(t,\tau,\bar{x}) + g'_{\rm m}(t,\tau,\bar{x})(x(\tau)-\bar{x}) \label{F_fb2_Supp}
\end{eqnarray}
where the first term in \eqn{F_fb1_Supp}, $g_{\rm m}(t,\tau,x)$, is a generalized gain kernel that allows an arbitrary coherent response at time $t$, dependent on the full prior history of the oscillator position $x$. The second term includes all incoherent forcing with $\xi_{\rm m}(\tau)$ being a unitless white-noise Wiener process and $n_{\rm m}(t,\tau,x)$ a memory kernel that permits non-Markovian, non-stationary and position dependent noise. The actuation force can be expressed similarly by replacing $x$ with $\tilde{x}$ in \eqn{F_fb1_Supp}. In general, both terms in \eqn{F_fb1_Supp} can introduce nonlinearities and instabilities which can give rise to detrimental effects like saturation and nonlinear dissipation~\cite{Harris12_PRA, Eichler11_NatNano}. Here, we wish to show that no advantage is possible from feedback in the case of a linear oscillator and linear feedback. Consequently, we Taylor expand both memory kernels $g_{\rm m}(t,\tau,x)$ and $n_{\rm m}(t,\tau,x)$ about the mean position of the oscillator $\bar{x}$ and neglect nonlinear terms. This gives $g_{\rm m}(t,\tau,x) = g_{\rm m}(t,\tau,\bar{x}) + g_{\rm m}'(t,\tau,\bar{x})(x-\bar{x})$ where $g_{\rm m}'(t,\tau,\bar{x})=\partial g_{\rm m}(t,\tau,x)/\partial x|_{x=\bar{x}}$ for the coherent memory kernel. Since the incoherent memory kernel is multiplied by the noise term $\xi_{\rm m}(\tau)$, nonlinearity occurs even in the second term of the expansion, so that for a linear oscillator $n_{\rm m}(t,\tau,x)= n_{\rm m}(t,\tau,\bar{x})$. Substituting these expressions into \eqn{F_fb1_Supp} results in \eqn{F_fb_Supp}, which can be further simplified into \eqn{F_fb2_Supp} by grouping all terms without dependence on the instantaneous displacement of the oscillator from its mean position into a single term $G_{\rm m}
(t,\tau,\bar{x})\!=\!g_{\rm m}(t,\tau,\bar{x}) \!+\! n_{\rm m}(t,\tau,\bar{x})\xi_{\rm m}(\tau)$ containing both noise and deterministic forcing.

The actuation force $F_{\rm act}$ can be similarly Taylor expanded into its component force as
\begin{eqnarray}
F_{\rm act}(t,\tilde{x}) &=& \int_{0}^{t}d\tau G_{\rm act}(t,\tau,\bar{x}) + g'_{\rm act}(t,\tau,\bar{x}) (\tilde{x}(\tau)-\bar{x}).
\end{eqnarray}
where $\tilde{x}(\tau)=x(\tau)-N(t,\bar{x})$. Here, the term $g'_{\rm act}(t,\tau,\bar{x})(x(\tau)-\bar{x})$ provides linear feedback modifying the susceptibility of the oscillator and driving its motion with white measurement noise $N(t,\bar{x})$. Taking the case of ideal feedback where no noise is introduced by the feedback loop itself, $G_{\rm act}(t,\tau,\bar{x})=g_{\rm act}(t,\tau,\bar{x})$, acts only to shift the mean position of the oscillator independent of the measurement record. Substituting the expressions for the actuation and combined forces into \eqn{eqmot_Supp} and Fourier transforming gives
\begin{eqnarray}
x(\Omega) \!\!&=&\!\! \chi(\Omega) \mathcal{F}_{\mathsmaller t \rightarrow \Omega} \Bigg\{\! \int_{0}^{t}\! d\tau G_{\rm m}(t,\tau,\bar{x}) \!+\! g'_{\rm m}(t,\tau,\bar{x}) (x(\tau)\!-\!\bar{x}) \Bigg. \nonumber \\ 
&&\hspace{-0.0cm} \Bigg.  + G_{\rm act}(t,\tau,\bar{x})+  g'_{\rm act}(t,\tau,\bar{x})(\tilde{x}(\tau)-\bar{x})  \Bigg\} \label{feed_Supp}
\end{eqnarray}
where $\mathcal{F}_{t\rightarrow \Omega}$ represents the Fourier transform and $\chi(\Omega)^{-1} \!=\! m\left[\Omega_m^2 - \Omega^2 + i \gamma \Omega\right]$ is the intrinsic mechanical susceptibility. It should be noted that the Fourier transform can only be taken if the oscillator is intrinsically stable. Consequently, scenarios such as regenerative amplification where feedback gives rise to exponential growth in displacement and eventually triggers a nonlinear response~\cite{Kippenberg05,Taylor12_OptExp} are excluded from the analysis here.

If feedback is not applied, both the linear feedback term and the feedback induced deterministic displacement are removed. In principle, the resulting change in mean oscillator displacement could modify the force sensitivity.  For example, in cavity optomechanics, the displacement could move the optical cavity toward resonance, modifying the optical power level in the cavity and through this the sensitivity. However, since the feedback induced displacement is deterministic and known, an equivalent displacement may be achieved without feedback by applying an external force. In the case of cavity optomechanics, operationally, this corresponds to ensuring the cavity is detuned from resonance equivalently in both the feedback and non-feedback scenarios at the beginning of the measurement. Applying this external force so that $\bar{x} = \bar{x}_{0}$, the oscillator position without feedback can then be simply obtained by omitting the feedback force $g'_{\rm act}(t,\tau,\bar{x})(x(\tau)-\bar{x})$ from \eqn{feed_Supp}
\begin{eqnarray}
x_{0}(\Omega) \!\!&=&\!\! \chi(\Omega) \mathcal{F}_{\mathsmaller t \rightarrow \Omega} \Bigg\{\!\! \int_{0}^{t}\!\! d\tau G_{\rm m}(t,\tau,\bar{x}) \!+\! g'_{\rm m}(t,\tau,\bar{x}) (x_{0}(\tau)-\bar{x}) \Bigg. \nonumber \\ 
&& \hspace{-.0cm} \Bigg.  \!+\! G_{ \rm act}(t,\tau,\bar{x}) \Bigg\}  \label{nofeed_Supp}
\end{eqnarray}
where the subscript $0$ is used to distinguish from the feedback case.

Subtracting \eqn{feed_Supp} from \eqn{nofeed_Supp} eliminates the common terms $G_{\rm act}(t,\tau,\bar{x})$ and $G_{\rm m}(t,\tau,\bar{x})$ and, after substituting $x(\Omega) = \tilde{x}(\Omega) - N(\Omega,\bar{x})$, results in a completely deterministic equation relating the time domain measurement record with and without feedback
\begin{eqnarray}
&&\hspace{-0.5cm} \tilde{x}(t) \!-\! \mathcal{F}^{-1}_{ \mathsmaller {\mathsmaller \Omega \rightarrow t}}\! \left\{ \chi(\Omega) \!\mathcal{F}_{ \mathsmaller {\mathsmaller t' \rightarrow \Omega}} \! \left\{ \!\int_{0}^{t'}\!\!\! d\tau \!\left[g'_{ \mathsmaller{\rm m}}(t'\!,\!\tau\!,\!\bar{x}) \!+\! g'_{\mathsmaller{\rm act}}(t'\!,\!\tau\!,\!\bar{x}) \right]\! (\tilde{x}(\tau)\!-\!\bar{x})\!  \right\}\!\! \right\}  \nonumber \\
&&=\tilde{x}_{0}(t) \!-\!\mathcal{F}^{-1}_{\mathsmaller \Omega \rightarrow t} \left\{ \chi(\Omega) \mathcal{F}_{ \mathsmaller t' \rightarrow \Omega} \! \left\{\! \int_{0}^{t'}\!\!\! d\tau g'_{\mathsmaller{\rm m}}(t'\!,\!\tau\!,\!\bar{x}) (\tilde{x}_{0}(\tau) \!-\! \bar{x}) \right\}\!\!\right\}
\label{equate_Supp}
\end{eqnarray}
where the dummy variable $t'$ has been introduced to distinguish the Fourier transform from its inverse. \eqn{equate_Supp} gives a non-trivial relationship between the measurement records with and without feedback in the presence of non-stationary forcing and feedback, non-Gaussian noise and correlations between measurement and process noise. However, for feedback and filtering to be equivalent, a causal filter must exist that maps $\tilde{x}_{0}\mapsto \tilde{x}$. To determine the existence of such a filter it is necessary solve for $\tilde{x}(\tau)$ as a function of $\tilde{x}_{0}(\tau)$ in \eqn{equate_Supp}.

Without loss of generality we may simplify \eqn{equate_Supp} by choosing our position coordinate such that the mean displacement
is zero $\bar{x} = 0$. The second term on the LHS of \eqn{equate_Supp} may then be expanded as
\begin{eqnarray}
&&\!\!\mathcal{F}^{-1}_{\Omega \rightarrow t} \left\{ \chi(\Omega) \mathcal{F}_{t' \rightarrow \Omega} \left\{ \int_{0}^{t'} \!\! d\tau \! \left[g'_{\rm m}(t',\tau,\bar{x})\!+\!g'_{\rm act}(t',\tau,\bar{x}) \right] \tilde{x}(\tau)  \right\}\! \right\}  \nonumber \\
	&&\hspace{-0.2cm} =\!\! \int_{-\infty}^{\infty}dt' \int_{0}^{t'}\!\! d\tau \left[g'_{\rm m}(t',\tau,\bar{x})\!+\!g'_{\rm act}(t',\tau,\bar{x}) \right] \tilde{x}(\tau) \chi(t-t')	\label{eq_1_Supp}\\
	&&\hspace{-0.2cm} =\!\! \int_{0}^{\infty}\!\! d\tau \left[\int_{-\infty}^{\infty}dt' \left[g'_{\rm m}(t',\tau,\bar{x}) \!+\! g'_{\rm act}(t',\tau,\bar{x}) \right] \chi(t-t')\right] \tilde{x}(\tau)	\label{eq_2_Supp} \\
	&&\hspace{-0.2cm} =\!\! \int_{0}^{\infty}\!\! d\tau \left(h_{\rm m}(t,\tau,\bar{x}) + h_{\rm act}(t,\tau,\bar{x})\right) \tilde{x}(\tau) \label{eq_3_Supp}
\end{eqnarray}
where the time shift property of Fourier transforms $\int_{-\infty}^{\infty}d\Omega e^{i\Omega (t-t')} \chi(\Omega)	=\chi(t-t')$ has been used to obtain \eqn{eq_1_Supp}. \eqn{eq_2_Supp} is found by enforcing causality, namely $g'_{j}(t',\tau)=0$ when $\tau>t'$, allowing the upper bound of the integral over $\tau$ to be extended to infinity and the order of integration to be rearranged. The term contained in the large square brackets of \eqn{eq_2_Supp} is a combined transfer function for the system which can be consolidated into terms denoted $h_{\rm m}(t,\tau,\bar{x})$ and $h_{\rm act}(t,\tau,\bar{x})$ that are zero when $\tau>t$, resulting in \eqn{eq_3_Supp}. Applying the same procedure to the RHS of \eqn{equate_Supp} simplifies the equation into a Fredholm equation of the second kind
\begin{eqnarray}
\tilde{x}(t)-\int_{0}^{\infty}d\tau \left(h_{\rm m}(t,\tau,\bar{x}) + h_{\rm act}(t,\tau,\bar{x})\right) \tilde{x}(\tau) \nonumber \\
=\tilde{x}_{0}(t)-\int_{0}^{\infty}d\tau h_{\rm m}(t,\tau,\bar{x}) \tilde{x}_{0}(\tau).
\label{equate2_Supp}
\end{eqnarray}
The kernel of the LHS is bounded and zero for $\tau >t$, so it is square integrable. Fredholm's theorem therefore guarantees the existence of solutions for $\tilde{x}$. In practice, without exploiting symmetries or assumptions about the kernel such equations are typically solved using numerical techniques~\cite{Polianin_Book, Jerri_Book}. Temporal discretization transforms \eqn{equate2_Supp} into matrix form
\begin{eqnarray}
&& \left[\mathbb{I}-\left(\mathbb{H}_{\rm m} + \mathbb{H}_{\rm act}\right)\right] 
\tilde{\bm{x}}		 = \left[\mathbb{I}-\mathbb{H}_{\rm m}\right]\tilde{\bm{x}}_{0}
\label{equate3}
\end{eqnarray}
where $\mathbb{I}$ is the identity matrix, $\tilde{\bm{x}}$ is a measurement record vector and the matrices $\mathbb{H}$ are the discretized kernels of \eqn{equate2_Supp}. The inner product between $\mathbb{H}$ and $\tilde{x}$ effects the integration. Consequently it is possible to solve for $\tilde{\bm{x}}$ to finally give
\begin{eqnarray}
\tilde{\bm{x}} = \left[\mathbb{I}-\left(\mathbb{H}_{\rm m} + \mathbb{H}_{\rm act} \right)\right]^{-1}\left[\mathbb{I} - \mathbb{H}_{\rm m} \right] \tilde{\bm{x}}_{0}
\end{eqnarray}
This expression shows that, for a linear oscillator, it is possible to exactly reproduce the measurement record that would be obtained with from a system with non-stationary processes by causally filtering the measurement record without feedback. This result is valid in both the quantum and classical regime and rigorously proves that no force sensing advantage is provided by linear feedback onto linear oscillators. This precludes enhancements of both sensitivity~\cite{Gavartin12_NatNano} and bandwidth~\cite{Miao12_NJP} beyond that achievable with estimation alone.




\end{document}